\documentclass[12pt]{article}
\usepackage{amssymb}
\usepackage{epsfig}

\catcode`\@=11
\def\numberbysection{\@addtoreset{equation}{section}
        \def\theequation{\thesection.\arabic{equation}}}

\def\beq{\begin{equation}}
\def\eeq{\end{equation}}
\numberbysection
\begin{document}
\begin{titlepage}
\begin{center}
\hfill \\
\vskip 1.in {\Large \bf Wilson loop in 2d noncommutative gauge theories} \vskip 0.5in P. Valtancoli
\\[.2in]
{\em Dipartimento di Fisica, Polo Scientifico Universit\'a di Firenze \\
and INFN, Sezione di Firenze (Italy)\\
Via G. Sansone 1, 50019 Sesto Fiorentino, Italy}
\end{center}
\vskip .5in
\begin{abstract}
We reconsider the perturbative expansion of the Wilson loop in 2d noncommutative gauge theories, using an improved integration method. For the class of maximally crossed diagrams in the $\theta \rightarrow \infty$ limit we find an intriguing formula, easily generalizable to all orders in perturbation theory.
\end{abstract}
\medskip
\end{titlepage}
\pagenumbering{arabic}
\section{Introduction}

In the recent years there has been a lot of interest in applying the methods of non-commutative geometry to quantum field theory and gravity. The principal aim of such investigations is to study how physics is modified by the extension of continuum space-time to non-commuting operators describing a non-local non-differentiable manifold, the type of
structure we expect to encounter at the Planck scale, where quantum gravity effects become relevant.

In this article we reconsider \cite{1}-\cite{3} the noncommutative $U(N)$ Yang-Mills theory, using as observable the Wilson loop, in a two-dimensional setting , where many exact non-perturbative results are known in the commutative case. Two dimensions are special since non-commutativity preserves Lorentz invariance, while in four dimensions the same property doesn't hold. However recent studies \cite{4}-\cite{5} have revealed the loss of invariance under area-preserving diffeomorphims. While classically the Wilson loop is independent of the shape of the contour, in the noncommutative gauge theory the loop correlators depend
on the path shape. Despite these difficulties we believe that in two dimensions non-perturbative results are obtainable, and we try to improve the perturbative analysis
of ref. \cite{1}-\cite{2} in order to reveal the integrability of the theory at its maximum extent.
We perform all the perturbative calculations without referring explicitly to a specific contour ( the case of the circle is already done in ref.\cite{3} with great details ), finding interesting results in complex coordinates.

Also at the commutative level, the light cone gauge is the simplest choice for calculating perturbatively the Wilson loop. Once that the gauge is fixed, two prescriptions for the gauge propagator are possible, the Cauchy principal value ( $PV$ ) and the Wu-Mandelstam-Leibbrandt ( $WML$ ) one.

At the non-commutative level we choose to work again with the light cone gauge and the $ WML $ prescription for the vector propagator hoping that the classical integrability of the Wilson loop is translated to the non-commutative setting.

By performing the first non-trivial order of the perturbative expansion, the only diagram which is affected by non-commutativity is the crossed one, where the $\theta$-dependent term is well defined and continuous in the classical limit. This behavior is typical of two dimensions, since in higher ones we expect to find singularities in the $\theta \rightarrow 0 $ limit. However the $\theta = 0$ remains somehow a singular value since it is reached in a non-analytic manner. In ref. \cite{3} it is shown that the continuity with $\theta = 0$ is reachable only in the $ WML $ prescription, while the $PV$ prescription, giving rise to an instantaneous ( point-like ) potential, is incompatible with non-commutativity of space-time.

The natural variable in which the perturbative expansion is analytic is $\frac{1}{\theta}$, and the $\theta \rightarrow \infty$ limit is another special point of interest, corresponding to maximal non-commutativity. Usually in higher dimensions one expect that non-commutativity makes the non-planar diagrams vanishing in the $\theta \rightarrow \infty$ limit, but in this case this property doesn't hold. The $\theta \rightarrow \infty$ limit is finite and well defined, and we will concentrate on this value to extend the notion of integrability of the Wilson loop at the non-commutative level.

We have completed the calculations of the next order and found that the only point of interest which is free of technical complications is the contribution from the maximally crossed diagrams in the $\theta \rightarrow \infty$ limit, where an intriguing formula is found, which can be easily generalized to all orders of perturbation theory. If our guess is true, this is the first example of a non-perturbative result for the Wilson loop in the non-commutative framework.

\section{Properties of NC quantum field theory}

A non-commutative quantum field theory can be described in two ways,i) replacing all the products among fields in the classical action with the star product, ii) using the operatorial formalism. Both definitions are totally equivalent and each one can be more useful for specific purposes. For example the star product formalism is well suited for perturbative studies, while the operatorial one is more helpful for searching soliton solutions or non-perturbative studies \cite{6}-\cite{7}. In this paper we will concentrate on the perturbative expansion of the Wilson loop, without concerning about the topological excitations \cite{8} which are far beyond our aim.

Perturbation theory is usually constructed from the quadratic part of the action, which defines the free propagator. As a first property, one can show that the free propagator of the $NC$ quantum field theory is the same as the commutative one, since the star product dependence reduces to a total derivative, which is trivial under suited boundary conditions. We will concentrate on the cases where this is possible.

Only the vertices of the theory are modified by the replacement with the star product, which changes dramatically the nature of the interaction. First, there is loss of invariance under arbitrary permutations of the momenta. Then it becomes crucial the distinction between planar and non-planar diagrams.

In planar diagrams, the $\theta$-dependent terms are strongly canceled and only an overall phase survives, depending only on the external momenta, which factorizes out in the loop integrals. The planar graphs of the $NC$ field theory have the same type of divergencies and singularities of the commutative ones \cite{9}.

Only the non-planar diagrams contains some non-commutative phases depending on the momenta running in the loops, modifying substantially the divergencies of the amplitude. The non-commutative phases act as a cutoff in the ultraviolet and in most cases non-planar graphs are ultraviolet convergent and free of divergencies. Therefore the introduction of a coordinate uncertainty principle is not enough to get rid of all ultraviolet divergencies.
We should also mention that such property is sensitive to the topology of the noncommutative space, since for example field theory defined on a non-commutative cylinder is ultraviolet finite as shown is ref. \cite{90}.

Moreover other types of divergencies come out as a result of the modification of the non-planar diagrams. They produce another kind of singularities, namely infrared ones that arise when non-planar loop diagrams are inserted in more complicated loop graphs. This phenomenon is called UV/IR mixing \cite{10}-\cite{11}, since the infrared singularity is introduced when one integrates very high momenta in the loops.

The phenomenon of UV/IR mixing makes impossible decoupling the infrared and the ultraviolet regimes, which is the basis of the renormalizability program of quantum field theory. The $NC$ quantum field theories are therefore more difficult to analyze beyond the first few loop orders.

Another important property of $NC$ quantum field theories is the loss of unitarity when non-commutativity is considered between space and time coordinates as firstly discussed in ref. \cite{110}. These theories have an infinite number of time derivatives and are non-local in time. In this situation the concept of Hamiltonian looses its meaning and unitarity is lost. Space-time non-commutativity is also in trouble with causality as discussed in \cite{111}. In this paper we deal with such a case at least in the Minkowskian formulation, but we first do the analytic continuation to the Euclidean version, which is safe. Moreover we take as an observable the Wilson loop which is not very sensible to all the quantization problems felt by the Green functions. We can anticipate that in spite the difficulties inherent to a Minkowskian formulation the result in the Euclidean version is a regular one. Moreover by performing a direct calculation in Minkowski space the Wilson loop is in agreement with the analytic continuation of the Euclidean case \cite{1}-\cite{3}, at least with the $WML$ propagator ( the Cauchy principal value prescription has problems with the analytic continuation ). A similar behaviour occurs at the level of the Green functions, which keep their functional dependence on the momenta of the Euclidean case, although the analytic continuation to Minkowskian variables changes dramatically their analytic properties.

\section{Observables of NC gauge theory}

From now on we will concentrate on particular quantum field theories which enjoy the property of gauge invariance. The $NC$ gauge theory can be easily generalized using the star product formalism. In this paper we will concentrate on the observables of $NC$ gauge theory, postponing the analysis of scattering amplitudes to a future research.

When looking for observables in $NC$ gauge theory one has to deal with the problem that no gauge invariant local operators exist. Since the $NC$ gauge group contains also translations, a gauge invariant operator must be invariant under space-time symmetries and cannot be local.

Consider for example the following relation

\beq exp [ i k x ] * f(x) * exp [ - i k x ] = f [ x + \theta k ] \eeq

This equation can have two interpretations. The first is that the factor $ exp [ i k x ]$ is the generator of global translations. Secondly, for a field transforming in the adjoint representation, the same equation is also a $NC$ gauge transformation.

Therefore translations are a subset of the gauge transformations, and a gauge invariant operator must be translation invariant and global.

It is possible to construct global gauge invariant objects carrying a momentum different from zero, like the open Wilson lines ( $OWL$ ) \cite{12}-\cite{13}-\cite{14}-\cite{15}-\cite{16}-\cite{17}:

\begin{eqnarray}
& \ & W ( k, C ) = \ \int d^d x \ Tr \ W(x,C) * exp [ i k x ] \nonumber \\
& \ & W ( x, C ) = \ P_{*} exp \ [ i g \int_C A_i ( x + \xi ) d \xi^i ]
\end{eqnarray}

Here $C$ is a generic open line in space-time such as $\xi^i (0) - \xi^i (1) = l^i$, and $P_{*}$ denotes the $NC$ path ordering along the line $\xi$. The trace $Tr$ is carried over the color indices of the internal $U(N)$ gauge theory.

The path ordering can be written as a formal power series

\begin{eqnarray}
W(x,C) & = & \sum_{n+0}^{\infty} \ ( ig )^n \int^1_0 ds_1 ... \int^1_{s_{n-1}} ds_n \
\dot{\xi}^{i_1} ( s_1 ) .... \dot{\xi}^{i_n} ( s_n ) \nonumber \\
& \ & A_{i_1} ( x + \xi ( s_1 ) ) * ... * A_{i_n} ( x + \xi ( s_n ) )
\end{eqnarray}

which provides the natural bridge to the correlators of these quantities in the corresponding quantum theory.

We still have to impose the gauge invariance of $W(k,C)$ , knowing the transformation rule for the local operators $W(x,C)$:

\beq W' (x,C) = U(x) * W(x,C) * U^{\dagger} ( x + l ) \eeq

To recover gauge invariance one has to impose the constraint

\beq l = \theta k \eeq

which is a manifestation of the UV/IR mixing.

A closed loop is possible for the particular value $ k = 0 $. Now the open lines are a complete set of gauge invariant operators, while the closed loop were the same in the commutative case.

We should mention the physical implication of $NC$ Wilson lines and loops for the Aharonov-Bohm effect as discussed in ref. \cite{170}-\cite{171}.

\section{Wilson loop: first order}

We recall that in the commutative case it is possible to compute exactly both the partition function and the Wilson loop on a two-dimensional manifold, including instantons contributions. The key symmetry behind such integrability is the invariance under area-preserving diffeomorphims \cite{18}.

In the non-commutative case till now no exact forms have been computed for the Wilson loop, and our aim is to investigate if the classical integrability can be extended in some form. The starting action is the Yang-Mills functional

\beq S = - \frac{1}{4} \ \int \ d^2x \ Tr ( F_{\mu\nu} * F^{\mu\nu} ) \label{41} \eeq

where the non-commutative field strength $F_{\mu\nu}$ is given by

\beq F_{\mu\nu} = \partial_\mu A_\nu - \partial_\nu A_\mu - i g ( A_\mu * A_\nu - A_\nu * A_\mu ) \eeq

and $A_\mu$ is a $U(N)$ connection.

The action (\ref{41}) is invariant under $U(N)$ non-commutative gauge transformations

\beq \delta_{\lambda} A_\mu = \partial_\mu \lambda - i g ( A_\mu * \lambda - \lambda * A_\mu ) \eeq.

Working in the Minkowskian formulation, we choose the light-cone gauge $ n^\mu A_\mu = A_{-} = 0$, where the constant vector $n^{\mu} = \frac{1}{\sqrt{2}} ( 1, -1 )$, by imposing it with a Lagrange multiplier. The Faddeev-Popov ghosts decouple in this case even in non-commutative theories \cite{19}.

Then the action becomes quadratic since the field strength has only one component $F_{-+} = \partial_{-} A_{+}$. The corresponding propagator in momentum space can be defined with two different prescriptions, the Cauchy principal value

\beq D_{++}^{PV} = \ i \  PV \ [ k_{-}^{-2} ] \eeq

and the Wu-Mandelstam-Leibbrandt ( $WML$ )

\beq D_{++}^{WML} = \ i \  {[ k_{-} + i \epsilon k_{+} ]}^{-2} \eeq

While the first one, when transformed to coordinate space, correspond to an instantaneous potential, which is incompatible with non-commutativity

\beq D_{++}^{PV} = \ -\frac{i}{2} \ | x_{+} | \delta(x_{-})  \eeq

the other not

\beq D_{++}^{WML} = \ - \frac{1}{2 \pi} \frac{ x_{+} }{( x_{-} - i \epsilon x_{+} ) } \eeq

Here $ x_{+} = x^{-} = \frac{1}{\sqrt{2}} ( x^0 - x^1 ) $ and $ x_{-} = x^{+} = \frac{1}{\sqrt{2}} ( x^0 + x^1 ) $.

The first prescription, used by 't Hooft \cite{20}, leads directly to the exact result for the Wilson loop, included topological excitations:

\beq W_{PV} = exp ( \frac{i}{2} g^2 N {\cal A} ) \eeq

whereas the $WML$ propagator \cite{21}-\cite{22}-\cite{23} gives rise to an exact resummation of all perturbative orders, without taking into account automatically topological effects:

\beq W_{WML} = \ \frac{1}{N} \ exp ( \frac{i}{2} g^2 {\cal A} ) \ L^{(1)}_{N-1} ( - i g^2 {\cal A}) \eeq
and $L^{(1)}_{N-1}$ is a Laguerre polynomial \cite{24}.

We are interested in the Euclidean formulation, which is obtained with the Wick rotation of the $WML$ propagator:

\beq D_{++}^{WML} = - \frac{2}{{( k_1 - i k_2 )}^2} \eeq

Accordingly, the Minkowski contour of integration is changed to a complex contour with the rule $ ( x_0, x_1) \rightarrow ( i x_0, x_1) $, and the area ${\cal A}$ is converted into
i ${\cal A}$. The Euclidean formulation cannot be defined with the $PV$ propagator.

Then, once quantized the theory, we can compute the vacuum expectation value of the non-commutative Wilson loop ( the case $k=0$ of the open Wilson line discussed in the previous paragraph )

\beq W[C] = \frac{1}{N} \ < \ \int d^2 x \ Tr \ {\cal T} \ P_{*} exp \left( ig \int_C A_{+} ( x + \xi(s) ) d \xi_{-} (s) \right) \ > \ \eeq

where $C$ is a closed contour parameterized by $\xi(s), 0 \leq s \leq 1$, and $P_{*}$ denotes the non-commutative path ordering along $x(s)$.

The closed Wilson loop $W[C]$ can be expanded as a power series in the coupling constant $g$:

\begin{eqnarray}
W[C] & = &  - \frac{1}{N} \sum_{n=0}^{\infty} ( - g^2 )^n \int^1_0 ds_1 ... \int^1_{s_{2n-1}} \ \dot{\xi}_{-} ( s_1 ) ... \dot{\xi}_{-} ( s_{2n} ) \nonumber \\
& \ & \int d^2 x < 0 | Tr \ {\cal T} \ [ \ A_{+} ( x + \xi(s_1)) * ... * A_{+} ( x + \xi(s_{2n})) \ ] \ | 0 > \end{eqnarray}

where each vacuum expectation value can be computed in momentum space

\begin{eqnarray}
& \ & \int d^2 x < 0 | \ Tr \ {\cal T} \ [ \ A_{+} ( x + \xi(s_1)) * ... * A_{+} ( x + \xi(s_{2n})) \ ] \ | 0 > = \nonumber \\
& \ & = Tr ( T^{a_1} ... T^{a_{2n}} ) \ \int d^2 p_1 ... \int d^2 p_{2n} \ exp [ i p_1 \xi(s_1) ] ... exp [ i p_{2n} \xi(s_{2n}) ] \nonumber \\
& \ & D_{++}^{a_1 ... a_{2n}} ( p_1 ... p_{2n} ) \ \int d^2 x \ exp [ i p_1 x ] * ... * exp [ i p_{2n} x ] \end{eqnarray}

The correlation function $D_{++}^{a_1 ... a_{2n}}$ between $2n$ fields $A_{\mu}^a$ is completely analogous to the commutative case. The Moyal deformation factor enter in the following formula

\beq \int d^2 x \ exp [ i p_1 x ] * ... * exp [ i p_{2n} x ] = \delta ( \sum_{i=1}^{2n} p_i ) exp \left[ - \frac{i}{2} \sum_{i<j} \tilde{p}_i  p_j \right] \eeq

where $ \tilde{p} q = p_{\mu} \theta^{\mu\nu} q_{\nu} $.

If we limit ourself to the first order contribution it is enough to consider

\beq W[C] = 1 + g^2 W_2 + g^4 W_4 + O(g^6) \eeq

The single exchange diagram $W_2$ is exactly as in the commutative case, since $\tilde{p}_1 p_1 = 0$. The first non-commutative contribution comes from $W_4$, and only when the two propagators cross as in the pairing $(13)(24)$. The other two pairings are planar diagrams and they do not feel non-commutativity.

The contribution from the crossed diagram reads

\begin{eqnarray}
 & \ &  W_4^{(cr)} = \int [ ds ] \ \dot{\overline{x}} ( s_1 ) ... \dot{\overline{x}} ( s_4 )
\ \int \frac{ dp d\overline{p}}{4 \pi^2 \overline{p}^2 }
\ \int \frac{ dq d\overline{q}}{4 \pi^2 \overline{q}^2 }
 \ exp \left[ -\frac{\theta}{2} ( p \overline{q} - \overline{p} q ) \right] \nonumber \\
& \ & exp \left[ \frac{i}{2} ( p ( \overline{x} (s_1) - \overline{x} (s_3) ) +
\overline{p} ( x(s_1) - x(s_3) ) ) \right] \nonumber \\
& \ & exp \left[ \frac{i}{2} ( q ( \overline{x} (s_2) - \overline{x} (s_4) ) +
\overline{q} ( x(s_2) - x(s_4) ) ) \right]
\end{eqnarray}

where we introduced the short notation

\beq \int [ds] = \int^1_0 ds_1 .... \int^1_{s_3} ds_4 \ \ \ \ \ \overline{x} ( s_1 ) = x_1 ( s_1 ) - i x_2 ( s_1 ) \ \ \ \ p = p_1 + i p_2 \eeq

and $Tr ( T^a_N T^b_N T^a_N T^b_N ) = N$ for the crossed diagram. Notice that the non-commutative parameter has been Wick rotated to the Euclidean version $\theta \rightarrow i \theta$ where $ \theta = \theta^{01} $.

Let us define the reduced contribution $\tilde{W}_4^{cr}$:

\beq W_4^{cr} = \int [ds]  \ \dot{\overline{x}} ( s_1 ) ... \dot{\overline{x}} ( s_4 ) \ \tilde{W}_4^{cr} \eeq

Instead of working with real variables \cite{1}-\cite{3}, we integrate $\tilde{W}_4^{cr}$ in the $p$ complex variable and obtain:

\beq \tilde{W}_4^{cr} \ = \ - \frac{1}{4\pi} \ \int \ \frac{ dq d\overline{q}}{4 \pi^2 \overline{q}^2 } \ \frac{ x(s_1) - x(s_3) - i \theta q}{ \overline{x}(s_1) - \overline{x}(s_3) + i \theta \overline{q} } \ exp \ \left[ \frac{i}{2} ( q ( \overline{x}(s_2) - \overline{x} (s_4 )) + \overline{q} ( x(s_2) - x(s_4)) ) \right]
\eeq

It is more comfortable to simplify the notations, introducing the auxiliary variables

\begin{eqnarray}
& \ & A = x(s_1) - x(s_3) \nonumber \\
& \ & B = x(s_2) - x(s_4)
\end{eqnarray}

Our trick is to decompose the $q$-integral into simpler pieces, which can be solved exactly. We notice in fact that the denominator can be simplified as

\begin{eqnarray} \frac{1}{\overline{q}^2} \ \frac{ A - i \theta q }{ \overline{A} + i \theta \overline{q}} & = & \left[ \frac{A}{\overline{A}} \ \frac{1}{\overline{q}^2} + i \theta \ \frac{1}{\overline{A}} \ \frac{\partial}{\partial \overline{q}} \ \left( \frac{q}{\overline{q}} \right) - \ \theta \ \frac{ A \overline{q}}{\overline{A}^2} \ \left ( \frac{1}{\overline{q}^2} \right) \right. \nonumber \\
& \ & -  \left. \frac{\theta^2}{\overline{A}^2} \ \frac{q}{\overline{q}} \ - \ \frac{\theta^2}{\overline{A}^2} \ \frac{A - i \theta q }{ \overline{A} + i \theta \overline{q}}  \ \right] \label{421}
\end{eqnarray}

giving rise to the following straightforward integration

\beq \tilde{W}_4^{cr} = \frac{1}{{(4 \pi)}^2} \left[ \frac{A \  B}{\overline{A} \ \overline{B}} \ + \ \frac{ 4 \theta^2 }{ \overline{A}^2 \ \overline{B}^2 } \left(
exp \left[ \frac{  A \overline{B} - \overline{A} B }{2\theta}  \right] - 1 - \frac{A \overline{B} - \overline{A} B}{2\theta} \right) \right] \label{422}
\eeq

This is our main result, which we have extended to the next order. In formula (\ref{422}) it is easy to check the continuity with the classical result:

\beq \tilde{W}_4^{cr} ( \theta = 0 ) = \frac{1}{{(4 \pi)}^2} \frac{A \  B}{\overline{A} \ \overline{B}} \eeq

while in the $\theta \rightarrow \infty$ limit we suddenly obtain

\beq \tilde{W}_4^{cr} ( \theta = \infty ) = \frac{1}{{(4 \pi)}^2} \left[\frac{1}{2} \left( \frac{A^2}{\overline{A}^2} + \frac{B^2}{\overline{B}^2} \right) \right] \label{424}\eeq

Generalization of this formula will be our next step. To achieve this we will consider the sixth order loop calculation.

\section{Wilson loop: higher orders}

We organize the sixth-order according to the topology of the diagrams. Diagrams without any crossing are not affected by the Moyal phase. Then there are six diagrams with a single crossing, three with double crossing and $W_{(14)(25)(36)}$, which is a maximally crossed diagram. We will work out all three cases, but we can anticipate that the more interesting diagram is the last one, where all propagators cross.

Let us start with an example of singly-crossed diagrams.

\begin{eqnarray}
W_{(16)(24)(35)} \ &  = & \ - N \ \int \ [ds] \ \dot{\overline{x}} ( s_1 ) \dot{\overline{x}} ( s_2 ) .... \dot{\overline{x}} ( s_6 ) \ \tilde{W}_{(16)(24)(35)} \nonumber \\
\tilde{W}_{(16)(24)(35)} \ &  = & \ \int \ \frac{dp d\overline{p}}{4 \pi^2 \overline{p}^2}
\ \int \ \frac{dq d\overline{q}}{4 \pi^2 \overline{q}^2} \ \int \ \frac{dk d\overline{k}}{4 \pi^2 \overline{k}^2} \ exp \left[ - \frac{\theta}{2} ( q \overline{k} - \overline{q} k ) \right] \nonumber \\
& \ & exp \left[ \frac{i}{2} ( p ( \overline{x} ( s_1 ) - \overline{x} ( s_6 )) + \overline{p} ( x ( s_1 ) - x ( s_6 ) )) \right] \nonumber \\
& \ & exp \left[ \frac{i}{2} ( q ( \overline{x} ( s_2 ) - \overline{x} ( s_4 )) + \overline{q} ( x ( s_2 ) - x ( s_4 ) )) \right] \nonumber \\
& \ & exp \left[ \frac{i}{2} ( k ( \overline{x} ( s_3 ) - \overline{x} ( s_5 )) + \overline{k} ( x ( s_3 ) - x ( s_5 ) )) \right]
\end{eqnarray}

It is not difficult to notice that

\beq \tilde{W}_{(16)(24)(35)} = - \frac{1}{4\pi} \ \frac{ x(s_1) - x(s_6) }{ \overline{x}(s_1) - \overline{x}(s_6) } \ \tilde{W}_{(24)(35)} \eeq

and this integral has been done in the previous paragraph (\ref{422}).

We then concentrate our discussion to a doubly-crossed diagram, for example

\begin{eqnarray}
\tilde{W}_{(15)(24)(36)} \ &  = & \ \int \ \frac{dp d\overline{p}}{4 \pi^2 \overline{p}^2}
\ \int \ \frac{dq d\overline{q}}{4 \pi^2 \overline{q}^2} \ \int \ \frac{dk d\overline{k}}{4 \pi^2 \overline{k}^2} \  \nonumber \\
& \ & exp \left[ \frac{i}{2} ( p ( \overline{x} ( s_1 ) - \overline{x} ( s_5 )) + \overline{p} ( x ( s_1 ) - x ( s_5 ) )) \right] \nonumber \\
& \ & exp \left[ \frac{i}{2} ( q ( \overline{x} ( s_2 ) - \overline{x} ( s_4 )) + \overline{q} ( x ( s_2 ) - x ( s_4 ) )) \right] \nonumber \\
& \ & exp \left[ \frac{i}{2} ( k ( \overline{x} ( s_3 ) - \overline{x} ( s_6 )) + \overline{k} ( x ( s_3 ) - x ( s_6 ) )) \right] \nonumber \\
& \ & exp \left[ - \frac{\theta}{2} ( p \overline{k} - \overline{p} k ) \right] \
exp \left[ - \frac{\theta}{2} ( q \overline{k} - \overline{q} k ) \right] \label{53}
\end{eqnarray}

By looking at the integral (\ref{53}), since both Moyal terms depend on $k$, it is more
convenient integrating firstly in the variables $p$ and $q$ to get:

\begin{eqnarray}
\tilde{W}_{(15)(24)(36)} \ &  = & \ \frac{1}{{(4\pi)}^2} \ \int \ \frac{dk d\overline{k}}{4 \pi^2 \overline{k}^2} \ \frac{ x(s_1) - x(s_5) - i \theta k }{ \overline{x}(s_1) - \overline{x}(s_5) + i \theta \overline{k} } \ \frac{ x(s_2) - x(s_4) - i \theta k }{ \overline{x}(s_2) - \overline{x}(s_4) + i \theta \overline{k} } \nonumber \\
& \ & exp \left[ \frac{i}{2} ( k ( \overline{x} ( s_3 ) - \overline{x} ( s_6 )) + \overline{k} ( x ( s_3 ) - x ( s_6 ) )) \right] \label{54}
\end{eqnarray}

A shorter notation is needed and we define the auxiliary variables

\begin{eqnarray}
& \ & A = x(s_1) - x(s_5) \nonumber \\
& \ & B = x(s_3) - x(s_6) \nonumber \\
& \ & C = x(s_2) - x(s_4)
\end{eqnarray}

so that the integral (\ref{54}) looks like

\beq
\tilde{W}_{(15)(24)(36)} \  =  \ \frac{1}{{(4\pi)}^2} \ \int \ \frac{dk d\overline{k}}{4 \pi^2 \overline{k}^2} \ \frac{ A - i \theta k }{ \overline{A} + i \theta \overline{k} } \ \frac{ C - i \theta k }{ \overline{C} + i \theta \overline{k} }
\  exp \left[ \frac{i}{2} ( k \overline{B} + \overline{k} B ) \right] \label{56}
\eeq

Eq. (\ref{56}) contains more factors than eq. (\ref{421}), but such complexity is not a big problem. Again the denominator can be decomposed as

\begin{eqnarray}
& \ & \frac{1}{\overline{k}^2} \ \frac{1}{ ( \overline{A} + i \theta \overline{k} )
( \overline{C} + i \theta \overline{k} ) } \ = \ \left[ \ \frac{1}{\overline{A}\overline{C}} \ \frac{1}{\overline{k}^2} \ - \ i \theta \frac{ \overline{A} + \overline{C}}{\overline{A}^2 \overline{C}^2} \ \frac{1}{\overline{k}} - \right. \nonumber \\
& \ & \left. - \theta^2 \ \frac{1}{\overline{A}^2 ( \overline{C} - \overline{A} )} \ \frac{1}{ \overline{A} + i \theta \overline{k}} \ + \ \theta^2 \ \frac{1}{\overline{C}^2 ( \overline{C} - \overline{A} )} \ \frac{1}{ \overline{C} + i \theta \overline{k}} \right]
\end{eqnarray}

from which the integrations are straightforward

\begin{eqnarray}
\tilde{W}_{(15)(24)(36)} \ & = &  \ - \frac{1}{{(4\pi)}^3} \left\{ \ \frac{ABC}{\overline{A}\overline{B}\overline{C}} \ + \ \frac{16 \theta^3}{\overline{A}^2 \overline{C}^2 ( \overline{C}-\overline{A} ) \overline{B}^3 } \left[ \right. \right. \nonumber \\
& \ & \overline{C}^2 \left( exp\left[ \frac{ A \overline{B} - \overline{A} B}{2\theta} \right] - 1 - \frac{ A \overline{B} - \overline{A} B}{2\theta} - \frac{ {( A \overline{B} - \overline{A} B )}^2 }{8 {\theta}^2} \right) \nonumber \\
& \ & \left. - \overline{A}^2 \left( exp\left[ \frac{ C \overline{B} - \overline{C} B}{2\theta} \right] - 1 - \frac{ C \overline{B} - \overline{C} B}{2\theta} - \frac{ {( C \overline{B} - \overline{C} B )}^2 }{8 {\theta}^2} \right) \right]\nonumber \\
& \ & - \frac{4 \theta^2 ( A-C )}{ \overline{A}^2 \overline{B}^2 \overline{C}^2 ( \overline{C} - \overline{A} ) } \ \left[ \
\overline{C}^2 \left( exp\left[ \frac{ A \overline{B} - \overline{A} B}{2\theta} \right] - 1 - \frac{ A \overline{B} - \overline{A} B}{2\theta} \right) +
\right. \nonumber \\
& \ & + \left. \left.\overline{A}^2 \left( exp\left[ \frac{ C \overline{B} - \overline{C} B}{2\theta} \right] - 1 - \frac{ C \overline{B} - \overline{C} B}{2\theta} \right) \right]
\right\} \end{eqnarray}

For $\theta \rightarrow 0$ we reobtain the classical limit

\beq \tilde{W}_{(15)(24)(36)} (\theta=0) \ = \ - \frac{1}{{(4\pi)}^3} \ \frac{ABC}{\overline{A}\overline{B}\overline{C}} \eeq

while for $\theta \rightarrow \infty$ we obtain the maximally noncommutative point:

\begin{eqnarray}
\tilde{W}_{(15)(24)(36)} (\theta=\infty) \ & = & \ - \frac{1}{{(4\pi)}^3} \ \left[ \
\frac{1}{3} \ \frac{ \overline{C}^2 A^3 - \overline{A}^2 C^3}{\overline{A}^2 \overline{C}^2 ( \overline{C}- \overline{A} ) } \ - \ \frac{1}{2} \ \frac{(A-C)(\overline{C}^2 A^2 + \overline{A}^2 C^2 )}{\overline{A}^2 \overline{C}^2 ( \overline{C}-\overline{A} ) } \right. \nonumber \\
& + & \left. \ \frac{1}{3} \frac{B^3}{\overline{B}^3}
\right]
\end{eqnarray}

Then we arrive at the last diagram, which is by far the more interesting. The maximally crossed diagram gives rise to

\begin{eqnarray}
\tilde{W}_{(14)(25)(36)} \ &  = & \ \int \ \frac{dp d\overline{p}}{4 \pi^2 \overline{p}^2}
\ \int \ \frac{dq d\overline{q}}{4 \pi^2 \overline{q}^2} \ \int \ \frac{dk d\overline{k}}{4 \pi^2 \overline{k}^2} \  \nonumber \\
& \ & exp \left[ \frac{i}{2} ( p ( \overline{x} ( s_1 ) - \overline{x} ( s_4 )) + \overline{p} ( x ( s_1 ) - x ( s_4 ) )) \right] \nonumber \\
& \ & exp \left[ \frac{i}{2} ( q ( \overline{x} ( s_2 ) - \overline{x} ( s_5 )) + \overline{q} ( x ( s_2 ) - x ( s_5 ) )) \right] \nonumber \\
& \ & exp \left[ \frac{i}{2} ( k ( \overline{x} ( s_3 ) - \overline{x} ( s_6 )) + \overline{k} ( x ( s_3 ) - x ( s_6 ) )) \right] \nonumber \\
& \ &
exp \left[ - \frac{\theta}{2} ( p \overline{q} - \overline{p} q ) \right] \
exp \left[ - \frac{\theta}{2} ( p \overline{k} - \overline{p} k ) \right] \
exp \left[ - \frac{\theta}{2} ( q \overline{k} - \overline{q} k ) \right]
\end{eqnarray}

In this case the Moyal terms are mixing all variables and the reduction to a single integration seems rather difficult. However, by defining

\begin{eqnarray}
& \ & A = x(s_1) - x(s_4) \nonumber \\
& \ & B = x(s_3) - x(s_6) \nonumber \\
& \ & C = x(s_2) - x(s_5)
\end{eqnarray}

we can rewrite the integral to compute

\begin{eqnarray}
\tilde{W}_{(14)(25)(36)} \ &  = & \ \int \ \frac{dp d\overline{p}}{4 \pi^2 \overline{p}^2}
\ \int \ \frac{dq d\overline{q}}{4 \pi^2 \overline{q}^2} \ \int \ \frac{dk d\overline{k}}{4 \pi^2 \overline{k}^2} \  exp \left[ - \frac{\theta}{2} ( p \overline{q} - \overline{p} q ) \right] \nonumber \\
& \ & exp \left[ \frac{i}{2} ( p ( \overline{A} + i \theta \overline{k} ) + \overline{p} ( A - i \theta k ) ) \right] \nonumber \\
& \ & exp \left[ \frac{i}{2} ( q ( \overline{C} + i \theta \overline{k} ) + \overline{q} ( C- i \theta k ) ) \right]  \ exp \left[ \frac{i}{2} ( k \overline{B} + \overline{k} B ) \right] \label{513}
\end{eqnarray}

into a form which suggests the two possible integrations, using our previous experience on the $g^4$ contribution. In fact, the double integral

\beq \int \ \frac{dp d\overline{p}}{4 \pi^2 \overline{p}^2}
\ \int \ \frac{dq d\overline{q}}{4 \pi^2 \overline{q}^2} \
exp \left[ \frac{i}{2} ( p \overline{W}  + \overline{p} W ) \right] \  exp \left[ \frac{i}{2} ( q \overline{V} + \overline{q} V ) \right] \ exp \left[ - \frac{\theta}{2} ( p \overline{q} - \overline{p} q ) \right] \eeq

with $ W= A-i\theta k $, $ V= C- i \theta k $ can be solved by

\beq \frac{1}{{(4\pi)}^2}\ \left[ \ \frac{W \ V}{\overline{W} \ \overline{V}} \ + \ \frac{ 4 \theta^2 }{ \overline{W}^2 \ \overline{V}^2 } \left(
exp \left[ \frac{  W \overline{V} - \overline{W} V }{2\theta}  \right] - 1 - \frac{W \overline{V} - \overline{W} V}{2\theta} \right) \ \right] \ \eeq

The first term of this expression gives rise to an integral of the type

\beq \int \ \frac{dk d\overline{k}}{4 \pi^2 \overline{k}^2} \ \frac{ A - i \theta k }{ \overline{A} + i \theta \overline{k} } \ \frac{ C - i \theta k }{ \overline{C} + i \theta \overline{k} }
\  exp \left[ \frac{i}{2} ( k \overline{B} + \overline{k} B ) \right]
\eeq

that is already solved ( see the double-crossing integral formula (\ref{56})).

The other term which is $O(\theta^2)$ can be written as ( since $ k = - 2i \frac{\partial}{\partial\overline{B}} $ ) the action of the following differential
operator

\begin{eqnarray}
& \ & \left\{ exp \left( \frac{A \overline{C} - \overline{A} C}{2\theta} \right) \
exp \left[ - ( \overline{C} - \overline{A} ) \frac{\partial}{\partial \overline{B}}
 -( C - A ) \frac{\partial}{\partial B } \right] - 1 - \right. \nonumber \\
& \ & - \left. \left( \frac{A \overline{C} - \overline{A} C}{2\theta} \right) + ( \overline{C} - \overline{A} ) \frac{\partial}{\partial \overline{B}}
 + ( C - A ) \frac{\partial}{\partial B }  \right\} f( A,B,C) \label{517}
\end{eqnarray}

on $f(A,B,C)$ defined as

\beq f(A,B,C) \ = \ 4 \theta^2 \ \int \ \frac{dk d\overline{k}}{4 \pi^2 \overline{k}^2} \
\frac{1}{  {( \overline{A} + i \theta \overline{k} )}^2 {( \overline{C} + i \theta \overline{k} )}^2 } \ exp \left[ \frac{i}{2} ( k \overline{B} + \overline{k} B ) \right]
\eeq

We can reduce the powers of the factors in the denominator, noticing that

\beq f(A,B,C) \ = \ 4 \theta^2 \ \frac{\partial}{\partial\overline{A}} \ \frac{\partial}{\partial\overline{C}} \ \tilde{f} ( A, B, C) \eeq

where $\tilde{f} ( A, B, C)$ can be exactly computed:

\begin{eqnarray}
\tilde{f} (A,B,C) \ & = & \ \int \ \frac{dk d\overline{k}}{4 \pi^2 \overline{k}^2} \
\frac{1}{  ( \overline{A} + i \theta \overline{k} ) ( \overline{C} + i \theta \overline{k} ) } \ exp \left[ \frac{i}{2} ( k \overline{B} + \overline{k} B ) \right] = \nonumber \\
& = & \frac{B}{\overline{A} \overline{B} \overline{C}} + \frac{2\theta}{\overline{A}^2 ( \overline{C}-\overline{A}) \overline{B} } \left( exp \left[ \frac{  A \overline{B} - \overline{A} B }{2\theta}  \right] - 1 \right) - \nonumber \\
 & \ & - \frac{2\theta}{\overline{C}^2 ( \overline{C}-\overline{A}) \overline{B} } \left( exp \left[ \frac{  C \overline{B} - \overline{C} B }{2\theta}  \right] - 1 \right)
\end{eqnarray}

Since we are interested in the $\theta \rightarrow \infty$, we can avoid the evaluation of eq. ( \ref{517} ), by observing that in this limit the whole integral (\ref{513}) reduces to

\beq \tilde{W}_{(14)(25)(36)} ( \theta = \infty ) = \lim_{\theta \rightarrow\infty} \frac{1}{2} \ \frac{1}{{(4\pi)}^2} \ \int \ \frac{dk d\overline{k}}{4 \pi^2 \overline{k}^2} \ \left(
\frac{W^2}{\overline{W}^2} + \frac{V^2}{\overline{V}^2}
\right) \ exp \left[ \frac{i}{2} ( k \overline{B} + \overline{k} B ) \right] \eeq

because the other terms are at least of order $\frac{1}{\theta}$.

Then it is enough to compute

\begin{eqnarray}
I \ & = & \ \int \ \frac{dk d\overline{k}}{4 \pi^2 \overline{k}^2} \ {\left(
\frac{A - i \theta k}{\overline{A} + i \theta \overline{k}}
\right)}^2 \ exp \left[ \frac{i}{2} ( k \overline{B} + \overline{k} B ) \right] = \nonumber \\
& = & \ - \frac{1}{4\pi} \ \left\{ \ \frac{A^2 B}{\overline{A}^2 \overline{B}} \ + \ \frac{ 8 \theta^2 B}{\overline{A}^2 \overline{B}^3 } \ \left( exp \left[ \frac{  A \overline{B} - \overline{A} B }{2\theta}  \right] - 1 - \frac{A \overline{B} - \overline{A} B}{2\theta} \right) + \right. \nonumber \\
& + & \left. \frac{32 \theta^3}{\overline{A}^3 \overline{B}^3 } \ \left( exp \left[ \frac{  A \overline{B} - \overline{A} B }{2\theta}  \right] - 1 - \frac{A \overline{B} - \overline{A} B}{2\theta}  - \frac{{(A\overline{B} - \overline{A} B)}^2}{8 \theta^2}
\right) \right\}
\end{eqnarray}

In the $\theta \rightarrow \infty $ limit this integral reduces to

\beq I(\theta=\infty) \ = \ - \frac{1}{4\pi} \ \left[ \ \frac{2}{3} \ \frac{A^3}{\overline{A}^3} \ + \
 \frac{1}{3} \ \frac{B^3}{\overline{B}^3} \ \right] \eeq

therefore we conclude that

\beq \tilde{W}_{(14)(25)(36)} ( \theta = \infty ) \ = \ - \frac{1}{{(4\pi)}^3} \ \left[ \
\frac{1}{3} \ \left( \ \frac{A^3}{\overline{A}^3}\ + \ \frac{B^3}{\overline{B}^3} \ + \ \frac{C^3}{\overline{C}^3} \ \right) \ \right] \label{524}
\eeq

By comparing the formulas (\ref{424}) and (\ref{524}) we can make the guess that for the generic maximally crossed diagram $\tilde{W}_{(1,n+1)(2,n+2)...(n,2n)} \ = \ \tilde{W}^{mcr}$
the contribution for $\theta \rightarrow 0$ is

\beq \tilde{W}^{mcr} ( \theta = 0 ) \ = \ k \frac{A_1 A_2 ... A_n}{\overline{A}_1 \overline{A}_2 ... \overline{A}_n} \eeq

with $k$ a known constant, while for $\theta \rightarrow \infty $

\beq \tilde{W}^{mcr} ( \theta = \infty ) \ = \ \frac{k}{n} \ \left[ \ {\left( \frac{A_1}{\overline{A}_1}\right)}^n \ + \  {\left( \frac{A_2}{\overline{A}_2}\right)}^n
\ + \ ... \ + \ {\left( \frac{A_n}{\overline{A}_n}\right)}^n \ \right] \eeq

where $A_i = x( s_i ) - x( s_{n+i} )$. This is the most simple new result one can hope to obtain from noncommutative diagrams.

\section{Conclusions}

Two dimensional $U(N)$ gauge theories are an important example of exactly solvable quantum field theory. In the commutative case closed Wilson loops have been computed by summing the perturbative series in the light cone gauge. With the Cauchy principal value prescription the contribution from crossed diagrams vanish and only planar diagrams survive. The result is a simple area exponentiation which takes into account the topological sector. With the $WML$ prescription crossed diagrams no longer vanish and the perturbative series no longer exhibits Abelian-like exponentiation for $N>1$, containing only the zero instanton sector of the theory.

In the non-commutative case only crossed diagrams are affected by the Moyal phase and it is not surprising that a sensible expression can be obtained for a closed Wilson loop only using the $WML$ propagator. The 't Hooft form of the free propagator is singular when continued to the Euclidean case, in presence of the Moyal phase.

The modification due to the Moyal phase is continuous with the classical limit and finite in the case of maximal non-commutativity. We have been able to simplify the calculations of ref. \cite{1}-\cite{3} including the $O(g^6)$ order, and we have found a special class of diagrams, the maximally crossed ones, where it is possible to guess an all-order perturbative result.

Of course, it would be very interesting to sum the whole perturbative series as it was done in ref. \cite{24} in the commutative case. But the difficulty is, in our opinion, to estimate the combinatorial expression for non-maximally crossed diagrams, which is beyond our ability. In any case we hope that our method may help in clarifying the integrability of the theory in the non-commutative case.

\end{document}